\documentclass[lettersizze,journal]{IEEEtran}
\usepackage{amsmath,amssymb,amsfonts}
\usepackage{algorithm}
\usepackage{array}
\usepackage[table, dvipsnames]{xcolor}
\usepackage[caption=false,font=footnotesize]{subfig}
\usepackage{textcomp}
\usepackage{stfloats}
\usepackage{url}
\usepackage{verbatim}
\usepackage{graphicx}
\usepackage{cite}
\usepackage[noend]{algpseudocode}
\usepackage{tikz}
\usepackage{footnote}
\usepackage{multirow}
\usepackage{tabulary}
\makesavenoteenv{algorithm}
\usepackage{lipsum}
\usepackage{booktabs}
\usepackage{soul}
\usepackage{tablefootnote}
\usepackage{colortbl}
\usepackage{float} % This is necessary for the [H] specifier
\begin{document}
\title{Modal-based prediction of power system frequency response and frequency nadir\\
\thanks{This work was supported in part by the National Science Foundation (NSF) under Grant No. 2044629, and in part by CURENT, which is an NSF Engineering Research Center funded by NSF and the Department of Energy under NSF Award EEC-1041877.}
\thanks{Francisco Zelaya-Arrazabal, Sebastian Martinez-Lizana, and Hector Pulgar-Painemal are with the Department of Electrical Engineering and Computer Science, University of Tennessee, Knoxville, TN 37996, USA. e-mail: fzelayaa@vols.utk.edu, smartinezlizana@ieee.org, hpulgar@utk.edu}}
\author{%
Francisco Zelaya-Arrazabal,~\IEEEmembership{Graduate Student Member,~IEEE,} 
Sebastian Martinez-Lizana,~\IEEEmembership{Graduate Student Member,~IEEE,} 
Héctor Pulgar-Painemal,~\IEEEmembership{Senior Member,~IEEE}%
\vspace{-2em}} % Reduce space here

\maketitle
\begin{abstract}
This paper introduces a novel approach for predicting system frequency response (SFR) and frequency nadir based on modal analysis. By decomposing the full system dynamic response, the method identifies dominant modes based on their participation in frequency behavior and derives a closed-form expression for the frequency trajectory. Unlike traditional approaches based on the Average System Frequency (ASF) model, this method captures the true system dynamics and avoids oversimplified representations. The dominant modes exhibit low sensitivity to system parameters, enabling robust and accurate estimations across diverse operating conditions. The proposed approach is tested on two benchmark systems as well as the Salvadoran transmission planning network, demonstrating its scalability, precision, and adaptability. This methodology represents a shift from observing a simplified average system frequency response to a more detailed analysis focusing on system dynamics.
\end{abstract}
\begin{IEEEkeywords}
Frequency nadir, frequency response, nadir estimation, primary frequency regulation, frequency deviation. 
\vspace{-0.2in}
\end{IEEEkeywords}
\section{Introduction}
\IEEEPARstart{F}{requency} stability is a cornerstone of power system reliability and is becoming increasingly critical as the grid evolves with the growing integration of distributed energy resources (DERs), and inverter-based resources (IBRs). These can introduce fast and significant power imbalances that challenge system analysis, protection, and control \cite{matevosyan2021}. In this context, accurately estimating the system frequency response (SFR) and the frequency nadir—the lowest frequency reached after a disturbance—is vital for grid stability. These estimates guide key operational/planning decisions, including frequency relay coordination, under-frequency load shedding, spinning reserve allocation, and renewable integration\cite{badesa2021conditions, shekari2015analytical, sigrist2012principles, ortiz2020practical}.

% Conventional approaches to frequency nadir prediction typically rely on detailed nonlinear simulations \cite{ortiz2020practical, zelaya2023supplementary}, which, while accurate, are impractical for large-scale planning or real-time applications that require many simulations.

Frequency nadir can be estimated through nonlinear simulation-based approaches \cite{ortiz2020practical, zelaya2023supplementary}. While accurate, these techniques are impractical for large-scale planning and real-time applications, requiring an excessive number of simulations. To improve computational efficiency, simplified analytical methods based on the average system frequency (ASF) model have been widely used \cite{chan1972dynamic, anderson1990low, aik2006general, egido2009maximum, yan2015frequency, chengwei2017minimum, shi2018analytical, liu2020analytical, niu2022analytical, dong2023unified}. By aggregating generators, these models simplify dynamic behavior in response to power imbalances, improving efficiency but overlooking key dynamics—such as heterogeneous governor/turbine models, frequency-dependent loads, IBR controls, and high-order control loops \cite{egido2009maximum, liu2020analytical, shi2018analytical}. Moreover, as the grid continues to evolve, these simplified models require frequent heuristic adjustments and structural redesigns to accommodate new technologies and emerging dynamics \cite{yan2015frequency,chengwei2017minimum,niu2022analytical}.

This paper proposes a novel method to predict the SFR and frequency nadir using a modal-based approach that directly relies on the full mathematical model of the power system, without adopting simplified or average system models. By decomposing the system's response into its modal components and retaining only a small subset of dominant modes, the method enables a closed-form analytical estimation of both the frequency trajectory and its nadir. This approach not only eliminates the need for computationally intensive time-domain simulations but also captures system-wide dynamics that are often lost in traditional ASF-based models. A key advantage of the proposed method is its robustness under evolving grid conditions, including the growing presence of IBRs. The dominant modes used for estimation exhibit low sensitivity to parameter variations and remain valid across a broad range of operating scenarios. Therefore, the method does not require custom adjustments to incorporate non-synchronous generation. Instead, it inherently captures the system-wide dynamic effects introduced by these resources, offering a robust and accurate framework as the grid's composition evolves. 

The main contributions of this paper are as follows.  First, it introduces a closed-form algebraic expression for the SFR, derived using the system eigenstructure. Second, it proposes a practical framework for predicting the frequency nadir without the need for simplified aggregated models. This is supported by a comprehensive evaluation using standard test systems and the Salvadoran transmission planning network. The paper also provides a detailed sensitivity analysis of the dominant modes that govern the SFR, along with a separate evaluation of the influence of IBRs on nadir estimation. This work offers a distinctive perspective into predicting frequency response and its nadir by taking advantage of full system dynamics rather than relying on aggregated models. The paper is structured as follows: Section \ref{Sec:SFR_models} reviews traditional SFR models and their role in frequency nadir prediction. Sections III and IV introduce the proposed modal-based method, including its analytical derivation and implementation framework. Section V provides the case studies. Section VI examines the role of IBRs in nadir estimation. Finally, the conclusions of this work are presented in Section VII.

% \textcolor{blue}{The main contributions of this paper are as follows. First, it introduces a closed-form algebraic expression for the SFR, derived using the system eigenstructure. Second, it proposes a practical framework for predicting the frequency nadir without relying on exhaustive nonlinear simulations or simplified aggregate models. Finally, the work offers a distinctive perspective by taking advantage of full system dynamics—rather than reduced-order models—to predict the frequency trajectory and frequency nadir, an approach that, to the best of the authors’ knowledge, has not been explored in the existing literature.}

% This paper is structured as follows: Section \ref{Sec:SFR_models} reviews traditional SFR models and their role in frequency nadir prediction. Sections III and IV introduce the proposed modal-based method, including its analytical derivation and implementation framework. Section V presents a comprehensive evaluation across two standard benchmark systems and the Salvadoran transmission planning network, along with a sensitivity analysis of the dominant modes that influence the SFR. Section VI examines the role of IBRs in nadir estimation, highlighting how the proposed method captures system-wide dynamics without added modeling complexity. The conclusions are summarized in Section VII.

\section{Background of frequency response prediction} 
\label{Sec:SFR_models}
\subsection{Average system frequency (ASF) model}
\begin{figure}[t]
    \centering
    \includegraphics[width=0.35\textwidth]{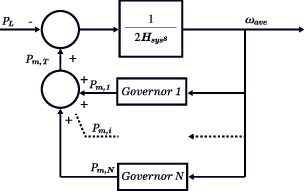} % 80% of the text width
    \caption{Average system frequency (ASF) model.}
    \label{fig:ASF_model}
\end{figure}

Assume the system has $N$ synchronous generators. This equivalent model aggregates all generators and describes the frequency response to load-generation imbalances through a single swing equation,
\begin{align}
\label{eq:ASF}
2H_{sys}\frac{d\omega_{avg}}{dt} &= P_{m,T} - P_{L}, \ \ \ 
\omega_{avg} &= \frac{\sum_{k=1}^N H_{k} \omega_{k}}{H_{sys}} 
\end{align}
where \(H_{sys} = \sum_{k=1}^N H_{k}\), \(P_{m,T} = \sum_{k=1}^N P_{m,k}\), and \(P_{L} = \sum_{k=1}^N P_{e,k}\) are the system's total inertia, compound mechanical power, and total electrical load (power output of generators), respectively. In addition, the turbine-governor loop is explicitly modeled for each machine, as described in Fig. \ref{fig:ASF_model}---reheat steam turbine-governor models are considered \cite{chan1972dynamic, kundur2007power}. The model considers the following assumptions: a) generation is dominated by reheat steam turbine generators, b) static loads are assumed—no voltage or frequency dependence; c) electromechanical oscillations are ignored; d) the effect of SGs' damper windings is neglected; e) all generators remain in synchronism during transients. Despite these simplifications, the model provides a good approximation of the SFR against power imbalances, but requires solving the entire system to obtain the frequency trajectory and to extract the nadir.

\begin{figure}[t]
    \centering
    \includegraphics[width=0.3\textwidth]{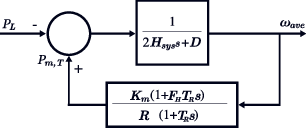} % 80% of the text width
    \caption{Low-order ASF model.}
    \label{fig:Low_order_ASF}
\end{figure}

To improve tractability and reduce complexity, this model was revised in \cite{anderson1990low} to reduce the combined turbine-governor dynamics into an aggregated single closed-loop representation. This simplified low-order ASF  model is depicted in Fig. \ref{fig:Low_order_ASF}. In addition, the damping factor \textcolor{black}{$D$} is now included, but its definition remains empirical. The parameter set $\Psi = [F_{H}, R, T_{R}]$ is assumed uniform across all machines, while $K_{m}$ reflects the system’s regulation capability. 
% This model remains widely adopted for benchmarking and as a simplified frequency representation in complex studies \cite{wang2019integrating,xie2025frequency}.

The ASF framework is extended to a multi-machine model to address its simplifying assumption that all generator prime movers are reheat-type steam turbines ~\cite{aik2006general,shi2018analytical}. This extension enables diverse turbine-governor types, such as gas and hydro, to be represented as equivalent reheat-type models by estimating the parameter set $\Psi$. The system's frequency nadir can be computed by identifying the time at which the derivative of the frequency deviation reaches zero~\cite{niu2022analytical}. While this generalization facilitates the aggregation of different machines into dominant models, it lacks a systematic method for determining $\Psi$, and the model’s accuracy depends heavily on the assumption that all turbine-governor types can be effectively approximated as reheat-type models.

\subsection{Semi-Analytical Modeling Methods}
Semi-analytical models aim to derive explicit expressions for computing the frequency nadir by simplifying the SFR behavior during the initial seconds following a power imbalance. Unlike the multi-machine ASF model, these methods employ low-order approximations (e.g., first-order transfer functions or polynomial fits) tailored to specific generation types.

\begin{figure}[t!]
    \centering
    \includegraphics[width=0.35\textwidth]{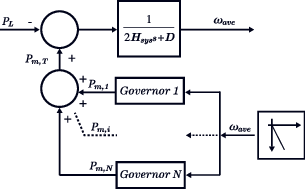} % 80% of the text width
    \caption{ASF with open loop.}
    \label{fig:ASF_BL}
\end{figure}

The SFR dynamics within the first 1--2 seconds are modeled by treating the governor input as a constant ramp, effectively decoupling the closed-loop control (Fig.~\ref{fig:ASF_BL}). Each turbine-governor model is approximated using first-order transfer functions derived from step-response fitting. Although closed-form expressions for the frequency nadir prediction can be obtained, it requires numerical methods to solve $N+1$ nonlinear equations \cite{egido2009maximum, chengwei2017minimum}. More advanced approaches introduce parabolic frequency decay and polynomial representations of governor response, enhancing accuracy over the first few seconds. However, this comes at the cost of increased numerical complexity and sensitivity to fitting procedures, as additional iterative steps are required to optimize the fitting window and polynomial order \cite{liu2020analytical}.

In summary, ASF-based methods, although practical, rely on a set of approximations that limit their accuracy and require solving the entire linearized model to estimate the frequency nadir. Semi-analytical methods introduce more assumptions that enable faster computation but reduce accuracy further. This paper proposes a novel approach that, instead of relying on ASF-based formulations, exploits the system's modal information to derive an explicit expression for estimating the SFR and frequency nadir. The proposed method provides a fast and accurate solution, without the need for aggregation or unnecessary reduced-order models for the primer movers.
 
\section{Modal-Based Frequency Response Estimation}
The proposed approach focuses on capturing the dynamic behavior of the system based on its eigenstructure. It leverages the fundamental principle that system responses can be characterized by the dominant dynamic modes that govern the time response of the state variables. By isolating and analyzing the most influential dynamic components, the method provides an efficient representation of system behavior while disregarding dynamics with negligible impact.

\subsection{Modal characterization of grid frequency response}
Consider a power system modeled through a set of differential-algebraic equations \cite{sauer2017power} such as $\dot{x} = f(x,y)$ and
$0 = g(x,y)$. Here, $x \in \mathbb{R}^n$, and $y \in \mathbb{R}^l$ are the vectors of state variables and algebraic variables, respectively. Assume the system is in steady-state at $x_0$ when a power imbalance occurs. This creates a frequency excursion and the system stabilizes at a new equilibrium point $x_e$. When linearized around $x_e$, the model becomes
$\Delta\dot{x} = J_{1}\Delta x + J_{2}\Delta y$ and $0 = J_{3}\Delta x + J_{4}\Delta y$. By eliminating algebraic equations through Kron reduction, the following simplified model is obtained,
\begin{equation}
\Delta \dot{x} = [J_{1} - J_{2}J_{4}^{-1}J_{3}]\Delta x  = A_s \Delta x
\end{equation}
with initial condition $\Delta x(0) = \Delta x_0 = x_0-x_e$, which is the difference between the pre- and post-disturbance states. To obtain the solution, the following similarity transformation is used: $V^{-1}A_sV = W^{T} A_sV = \Lambda$, where $V =\{v_i\}\in \mathbb{C}^{n \times n}$ is the right eigenvectors matrix of $A_s$, $W =\{w_i\}\in \mathbb{C}^{n \times n}$ is the left eigenvectors matrix of  $A_s$, $\Lambda = \text{diag} \{\lambda_i\} \in \mathbb{C}^{n \times n}$ is a diagonal matrix containing the eigenvalues of $A_s$, and the index $i \in \mathcal{I}=\{1,2,...,n\}$. Note that the set of right and left eigenvectors are considered to be orthonormal, i.e., $\langle w_i,v_k\rangle=w_i^Tv_k=\delta_{ik}~\forall~ i,k \in \mathcal{I}$, thus, $W^T=V^{-1}$. For an explicit solution of $\Delta \dot{x} = A_s\Delta x$, the transformed vector of state variables is defined as $q =V^{-1} \Delta x$ to obtain:
\begin{align}
\Delta \dot{x} &= A_s \Delta x,~~\Delta x(t_0)=\Delta x_0 \\
\Rightarrow V^{-1} \Delta \dot{x} &= V^{-1} A_s V q ,~~V^{-1}\Delta x(t_0)=V^{-1}\Delta x_0\\
\Rightarrow \dot{q} &= \Lambda q,~~q(t_0)=W^T\Delta x_0=q_0 \label{eq:Transf_model}
\end{align}
The explicit solution of Eq. \eqref{eq:Transf_model} is given by $q(t) = e^{\Lambda t} q_0$, where $q_0=W^T \Delta x_0=[w_1,...,w_n]^T \Delta x_0=[\langle w_1,\Delta x_0\rangle,...,\langle w_n,\Delta x_0\rangle]^T$. As $e^{\Lambda t}$ is a diagonal matrix, the solution of Eq. \eqref{eq:Transf_model} can be rewritten as, 
\begin{equation}
\label{Eq_decoupled}
q(t) = 
\begin{bmatrix}
e^{\lambda_1 t}\langle w_1, \Delta x_0\rangle \\
\vdots \\
e^{\lambda_n t}\langle w_n, \Delta x_0\rangle
\end{bmatrix},~~\forall~ t \ge t_0
\end{equation}
As a result, the solution for each transformed state variable $i$ becomes $q_i(t) = e^{\lambda_i t} \langle w_i, \Delta x_0\rangle,~ \forall~ i \in \mathcal{I}$. Given that $\Delta x = Vq=\sum_{i \in \mathcal{I}} v_iq_i$, the vector of state variables can be explicitly defined as $\Delta x(t) = \sum _{i \in \mathcal{I}} e^{\lambda_{i} t} \langle w_{i},\Delta x_0\rangle v_{i}$. Note that on the right hand side of this equation, all terms are scalars except for $v_i\in \mathbb{C}^n$. Finally, the explicit solution for a single state variable $k$ corresponds to:
\begin{equation}
\Delta x_{k}(t) = \sum _{i \in \mathcal{I}} e^{\lambda_{i} t} \langle w_i,\Delta x_0\rangle v_{i,k}  \  , \  \ \  \forall~ t \geq 0
\end{equation}
where $v_{i,k}$ is the $k^{th}$ term of the $i^{th}$ right eigenvector.

Let $\mathcal{Z} \subset \mathcal{I}$ be the index set related to the speed of the generators. Let $H_z$ be the inertia of the generator whose speed index is $z \in \mathcal{Z}$. The total inertia is $H_t=\sum_{z \in \mathcal{Z}} H_z$, and the frequency of the center of inertia (COI) becomes,
\begin{equation}
\label{wcoi}
\Delta \omega_{coi} = \sum\limits_{z \in \mathcal{Z}} C_z \Delta x_z=
\sum\limits_{z \in \mathcal{Z}} C_z
\sum _{i \in \mathcal{I}} e^{\lambda_{i} t} \langle w_i,\Delta x_0\rangle  v_{i,z}
\end{equation}
where $C_z = H_z /H_t$. This is an explicit and general expression for the frequency response of the system under any initial condition. In theory, this formulation is based on the linearized system around the post-disturbance equilibrium point, $x_e$. However, the eigenvalues $\lambda_i$, and eigenvectors $v_i$ and $w_i$, $~\forall~i\in \mathcal{I}$, are not highly sensitive to the power imbalance caused by the disturbances---this is quantitatively confirmed through the sensitivity analysis in Section~\ref{subsec:NE}. Thus, for the calculations of eigenvalues and eigenvectors in \eqref{wcoi}, the system is linearized around the pre-disturbance equilibrium point, $x_0$.

% \textcolor{red}{While the formulation above is based on initial condition perturbation theory \cite{hespanha2018linear}, this work adopts a practical variant suited for power system applications where the post-disturbance equilibrium \(x_e\) is not known a priori. Instead, the system is linearized around the pre-disturbance steady state \(x_0\), based on the fact that the dominant modes governing the SFR remain largely invariant under typical disturbances and operating condtions. This enables the estimation of \(x_e\) as a function of \(\Delta P\), as outlined in Section~III.C, and supports accurate SFR and frequency nadir predictions. The validity of this assumption is quantitatively confirmed through the sensitivity analysis in Section \ref{subsec:NE}.}

\subsection{Participation factor-guided mode selection}
\label{subsection:participation-factor}
As we are interested in the frequency excursion after a power imbalance, note that only those modes associated with the slow dynamic response of the rotating masses and governors are required for an accurate approximation. Let $\mathcal{M} \subset \mathcal{I}$ be the index set of the most influential modes in the frequency excursion, which will hereafter be referred to as \textit{frequency control modes}. The subset $\mathcal{M}$ is obtained as follows:
\begin{itemize}
    \item The participation factor matrix $P \in \mathbb{R}^{n \times n}$ is obtained as the element-wise multiplication of the left and right eigenvector matrices, i.e., $P= V \circ W$. 
    \item Let $\mathcal{S} \subset \mathcal{I}$ be the subset of the indices of state variables corresponding to rotor frequency and turbine-governor variables. Define  $P_\mathcal{S} \in \mathbb{R}^{s \times n}$, $ s= dim(\mathcal{S})$, as the matrix containing the participation factors for the $\mathcal{S}$-indexed state variables. 
    \item Define $\epsilon \in \mathbb{R}$ as a threshold, specific to each system, and obtained through trial and error during the initial run.
     \item The subset $\mathcal{M} \subset \mathcal{I}$ is then defined as the set of indices corresponding to those columns of $P_\mathcal{S}$ where all entries in the column are above the threshold \( \epsilon \), i.e., $\mathcal{M} = \{j \in \mathcal{I} \mid \forall i \in \mathcal{S}, \, |P_{\mathcal{S}_{ij}}| > \epsilon \}$
\end{itemize}
Based on these definitions, the inner product $\langle w_m,\Delta x_0\rangle$ can be estimated only with the components included in the subset $\mathcal{S}$, i.e., $\langle w_m,\Delta x_0\rangle\approx  \langle w_{m,\mathcal{S}},\Delta x_{0,\mathcal{S}}\rangle = \sum_{s\in \mathcal{S}} \omega_{m,s}\Delta x_{0,s}$ Thus, the frequency of the COI can be approximated as,
\begin{equation}
\label{eq:w_coi_approx}
\Delta \omega_{coi} \approx
\sum\limits_{z \in \mathcal{Z}}
\sum\limits_{m \in \mathcal{M}} C_z e^{\lambda_{m} t}
\langle w_{m,\mathcal{S}},\Delta x_{0,\mathcal{S}}\rangle v_{m,z}
\end{equation}

\subsection{$\Delta x_{0,\mathcal{S}}$ computation as a function of $\Delta P$}
\label{subsection: init con}
Each component of the vector $\Delta x_{0,\mathcal{S}}$ is defined as $\Delta x_{0,s}=x_{0,s} - x_{e,s},~\forall s \in \mathcal{S}$, where $x_{e,s}$ must be estimated based on the power imbalance, $\Delta P$, caused by the disturbance. These variables include the new steady-state frequency and the turbine-governor variables. Firstly, the post-disturbance steady-state frequency is obtained as \cite{kundur2007power}, 
\begin{equation}
    \Delta \omega=-\Delta P/\sum_{z \in \mathcal{Z}} \tfrac{1}{R_{D,z}}
    \label{eq:Post_freq}
\end{equation}
where $1/R_{D,z}$ is the speed regulation constant of machine $z$. As the steady-state frequency is determined, the turbine-governor models for all machines are decoupled. Given the steady-state frequency, for each turbine-governor model, use its set of DAEs, set the derivatives to zero, and solve simultaneously to determined the turbine-governor variables. 
For instance, using the IEESGO turbine-governor model \cite{zelaya2023supplementary} with a significantly large maximum power limit, and assuming $n_g$ generators, the post-disturbance turbine-governor variables are estimated as:
\begin{align}
y_{1,i}&=y_{2,i}=y_{3,i} = - \Delta \omega/R_{D,i}~~\forall~i \in \{1,2,...,n_g\}\\
T_{m,i}&=P_{C,i}- \Delta \omega /R_{D,i}~~\forall~i \in \{1,2,...,n_g\}
\end{align}
A similar procedure is applied for any turbine-governor model.

\section{Frequency Nadir prediction}
The frequency nadir can be found through a numerical procedure that minimizes $\Delta \omega_{coi}$ given by Eq. \eqref{eq:w_coi_approx}. However, an analytical approach can simplify and expedite this prediction. For instance, consider that only three modes are involved in the frequency excursion dynamics, the set of frequency control modes is formed by a complex conjugate pair of eigenvalues, $\lambda_{c1} = \lambda_{c2}^* \in \mathbb{C}$, and a real eigenvalue, $\lambda_r \in \mathbb{R}$. Thus, the frequency response estimation becomes:
\begin{equation}
\label{eq:1conjugated_1real}
    \Delta \omega_{coi} \approx e^{\lambda_{c_1} t} \gamma_{c_1} + e ^{\lambda_{c_2} t} \gamma_{c_2} + e ^{\lambda_{r} t} \gamma_{r}\ 
\end{equation}
where $\gamma_{m} = \sum_{z \in \mathcal{Z}} C_z \langle w_{m,\mathcal{S}}, \Delta x_{0,\mathcal{S}} \rangle v_{m,z},~\forall~m \in \mathcal{M}= \{c1, c2, r\}$. Given the conjugated nature of the first two terms on the right-hand side, these can be compactly expressed in terms of the cosine function $2e^{\alpha t} r_c cos(\beta t + \theta)$, where $\lambda_{c_1}=\alpha + j \beta $, $\gamma_{c_1}=E+jF$, $j=\sqrt{-1}$, $r_c = \sqrt{E^2 + F^2}$, and $\theta = \arctan{(F/E)}$. Thus, $\Delta \omega_{coi}$ and its derivative become,
\begin{align}
    \Delta \omega_{coi} &\approx 2e^{\alpha t} r_c cos(\beta t + \theta) + e ^{\lambda_{r} t} \gamma_{r}\label{eq:1conjugated_1real}\\
    \frac{d \Delta \omega_{coi}}{dt} &= R e^{\alpha t} cos(\beta t + \theta + \phi ) + \lambda_{r} e ^{\lambda_{r} t} \gamma_{r}\label{eq:derivativewcoi}
\end{align}
with $R = 2 r_c \sqrt{\alpha^2 + \beta ^2}$ and $\phi = \arctan{(\beta/\alpha)}$. The time occurrence of the frequency nadir corresponds to the time when Eq. \eqref{eq:derivativewcoi} vanishes. Note that this equation involves two transcendental functions, making it challenging to solve it analytically. Thus, a second-order Taylor expansion around an initial guess, $\tau$, can be used to approximate the right-hand side of Eq. \eqref{eq:derivativewcoi}, resulting in the following expression,
\begin{align}
\label{eq:wcoi_der}
     \frac{d \Delta\omega_{\text{coi}}}{dt} =  \left(a_1 t^2 + a_2 t + a_3 \right) + \left( a_4 t^2 + a_5 t + a_6 \right) = 0 
\end{align}
See appendix for the coefficients definition. Therefore, the estimated time of occurrence of the frequency nadir (hereafter referred to as nadir time) is:
\begin{equation}
\label{eq:tnadir}
    t_{nadir} = \frac{-(a_2+a_5) + \sqrt{(a_2+a_5)^2 - 4(a_1+a_4)(a_3+a_6)}}{2(a_1+a_4)}
\end{equation}
and the per unit frequency nadir prediction can be obtained through direct evaluation as $f_{nadir} = f_{e} + \Delta \omega_{coi} (t_{nadir})$, where $f_{e}$ is the post-disturbance steady-state frequency and can be extracted from Eq. \ref{eq:Post_freq}. If additional modes are required, either complex conjugated or real, the number of terms in Eqs. \eqref{eq:1conjugated_1real}-\eqref{eq:wcoi_der} will be increased; still, the solution for $t_{nadir}$ will be always the solution of a quadratic function. In Section \ref{sec:case study}, it will be shown that a small subset of modes is sufficient to accurately estimate the frequency response and its nadir.

\begin{figure}[t]
    \centering
    \includegraphics[width=0.48\textwidth]{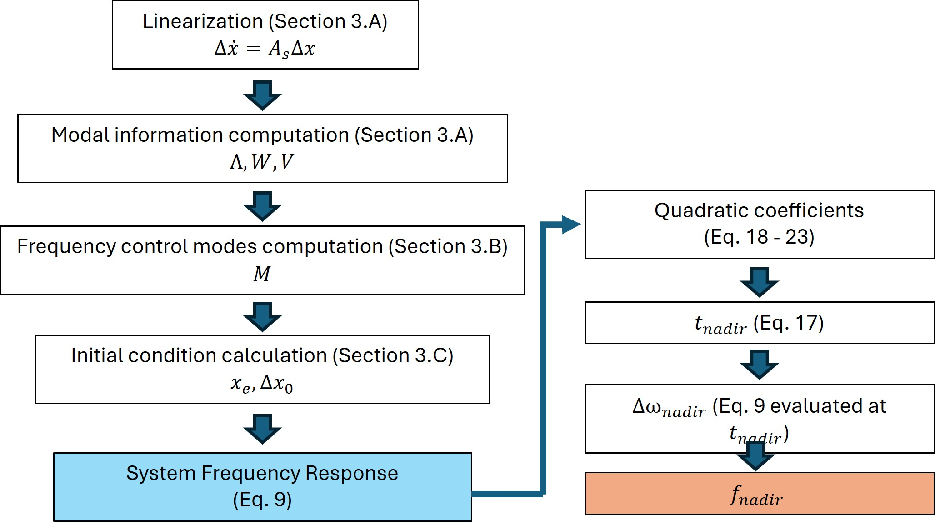} % 80% of the text width
    \caption{Modal-based SFR and nadir prediction flow chart.}
    \label{fig:SFR_flowchart}
\end{figure}

Fig. \ref{fig:SFR_flowchart} summarizes the steps to obtain the SFR and frequency nadir. Unlike other methods, the modal-based approach avoids model oversimplification and iterative solutions. This process leads to the algebraic representation of Eq. \ref{eq:w_coi_approx}, which is derived directly from the full dynamic model. Moreover, $t_{nadir}$ and $f_{nadir}$ can be easily calculated using Eq. \ref{eq:tnadir}. 

\section{Case Studies}
\label{sec:case study}
The case studies are based on two benchmark systems: the WSCC 9-bus test system and the New England 39-bus test system. The analysis is carried out using the Power System Simulator (PSSim), a MATLAB-based tool for power system analysis \cite{MartinezLizana2025}. The method is then further validated on El Salvador’s transmission planning system using the commercial PowerFactory software.  

% This section showcases the robustness of the modal-based approach, validated across these networks to demonstrate its scalability, precision, and adaptability.

\subsection{WSCC 9-bus test system}
The system consists of three SGs modeled as two-axis machines with IEEE Type-1 exciters and three types of governors. G1 and G2 are steam units (TGOV1 and IEESGO), while G3 is a gas unit (GAST). The governor parameters are presented in Appendix \ref{tab:governor_params}, and the system data is available in \cite{chow2020power}.

The initial evaluation considers three load steps—$3\%$, $6\%$, and $10\%$ of the total system load—occurring at $t = 1s$.  Table \ref{tab:comparison2} compares the frequency nadir, nadir time, and absolute errors of the modal-based approach with those from \cite{egido2009maximum} and \cite{liu2020analytical}, using time-domain simulation results as the benchmark.

\begin{table}[t]
    \centering
    \caption{Comparison of WSCC system nadir and time estimation across different methods}
    \renewcommand{\arraystretch}{1.1}
    \setlength{\tabcolsep}{4pt} 
    % \footnotesize % Fit table into one column
     \scriptsize
    \begin{tabular}{c|cc|cc|cc}
        \toprule
        \textbf{Case} & \multicolumn{2}{c|}{\textbf{3\%}} & \multicolumn{2}{c|}{\textbf{6\%}} & \multicolumn{2}{c}{\textbf{10\%}} \\
        & Nadir & Timing & Nadir & Timing & Nadir & Timing \\
        & (Hz) & (s) & (Hz) & (s) & (Hz) & (s) \\
        \midrule
        \rowcolor{green!20} \textbf{Benchmark} & 59.888 & 4.496 & 59.778 & 4.491 & 59.626 & 4.487 \\
        \midrule
        \rowcolor{Apricot!}\textbf{Modal-based} & 59.888 & 4.536 & 59.779 & 4.531 & 59.626 & 4.515 \\
        \textbf{\cite{egido2009maximum}}   & 59.911 & 3.431 & 59.823 & 3.431 & 59.702 & 3.431 \\
        \textbf{\cite{liu2020analytical}}  & 58.310 & 14.300 & 56.660 & 14.300 & 54.370 & 14.300 \\
        \midrule
        \rowcolor{gray!20} \multicolumn{7}{c}{\textbf{Absolute Error Comparison}} \\
        \midrule
       \rowcolor{Apricot!} \textbf{Modal-based} & 0.0002 & 0.0395 & 0.0003 & 0.04 & 0.0004 & 0.0278 \\
        \textbf{\cite{egido2009maximum}}       & 0.0228 & 1.0648 & 0.0451 & 1.0599 & 0.0766 & 1.0559 \\
        \textbf{\cite{liu2020analytical}}      & 1.5778 & 9.804 & 3.1183 & 9.8089 & 5.2555 & 9.8129 \\
        \bottomrule
    \end{tabular}
    \label{tab:comparison2}
\end{table}

\begin{figure}[t]
    \centering
    \hspace{-0.9cm} % Adjust this value as needed
    \includegraphics[width=0.53\textwidth]{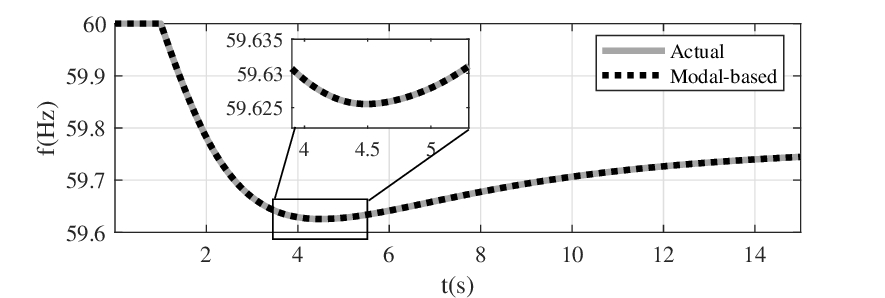}
    \caption{WSCC frequency after $10\%$ load step event.}
    \label{fig:10_per_event}
    \vspace{0.3cm}
    \hspace{-0.9cm} % Adjust this value as needed
    \includegraphics[width=0.53\textwidth]{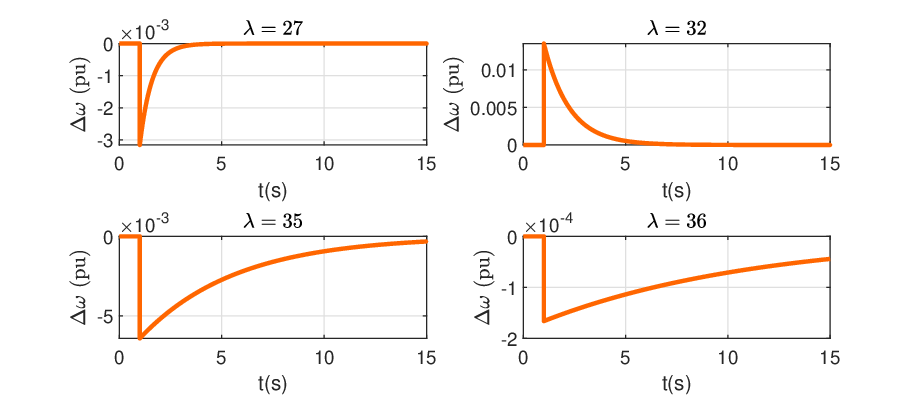}
    \caption{WSCC frequency control modes ($10\%$ load step).}
    \label{fig:nadir_dynamics}
\end{figure}

 The modal-based approach outperforms both methods in all scenarios. Method \cite{egido2009maximum} provides a simple analytical expression for nadir with reasonable accuracy but fails to estimate nadir time precisely. Method \cite{liu2020analytical} struggles with both nadir magnitude and time, likely due to its simplified form, in which the governor response is approximated as a second-order polynomial and the frequency response as a parabolic function. To obtain more accurate results, an iterative process is necessary to determine the optimal fitting time and polynomial. Moreover, a common drawback across these methods and others found in the literature is the difficulty in achieving a low error for nadir time. In contrast, the modal-based approach maintains a small error, thereby providing more reliable results. Fig. \ref{fig:10_per_event} compares the actual system frequency response with the modal-based response when there is a $10\%$ load step.
 
The modal-based prediction is determined by four dynamics corresponding to the real eigenvalues $\lambda_{\mathcal{M}} = \{\lambda_{27},\lambda_{32},\lambda_{35},\lambda_{36}\} = \{-1.679, -0.805, -0.213, -0.094\}$. These were selected using the participation factor-guided mode selection using a threshold of $\epsilon = 0.0001$. Notably, although the system comprises 38 state variables, only $\lambda_{\mathcal{M}}$ is required for accurate estimation. Fig. \ref{fig:nadir_dynamics} illustrates these four distinctive dynamics as $\gamma_{i}e^{\lambda_{i}t}$.

Finally, in terms of time performance, using a personal computer with an Intel(R) Core(TM) i7-9750H CPU and 16 GB of RAM—and assuming that the modal information has been previously computed—the average execution time for the frequency nadir and time estimation is 0.0115 seconds, whereas comparison methods report 0.076 seconds \cite{egido2009maximum} and 0.021 seconds \cite{liu2020analytical}, respectively, for the same system.

% \vspace{-0.5cm}
\subsection{New England 39-bus test system}
\label{subsec:NE}
The New England test system includes 10 two-axis machines, IEEE Type-1 exciters, and three governor types: TGOV1, IEESGO, and GAST. System parameters are provided in \cite{chow2020power}. Typical governor parameters were taken from \cite{neplan2015turbine} and are listed in Appendix \ref{tab:governor_params}. To observe a larger nadir, the H-constant of inertia for all generators was reduced by half. The system model has 136 state variables, and the proposed estimation requires only 4 eigenvalues $\lambda_{\mathcal{M}} = \{\lambda_{97},\lambda_{127},\lambda_{128},\lambda_{130}\} = \{-2.742, -0.191+0.111i, -0.191-0.111i, -0.093\}$, selected with $\epsilon = 0.0001$.

Table \ref{tab:comparisonGenTrip} presents the evaluation of two generator trip events: (1) G1 outage, resulting in a loss of $4\%$ of total system generation, and (2) G6 outage, resulting in a loss of $10.32\%$. The results show a low absolute error: $10$ mHz and $0.25$ s for frequency nadir and time, respectively. In terms of time performance, the average execution time for the predictions presented in this subsection is $0.01$ seconds. 
% The accuracy of the estimation is measured by the absolute percentage error (APE), defined as \( APE = \left| \frac{\text{actual value} - \text{estimated value}}{\text{actual value}} \right| \times 100\% \). 

% Table \ref{tab:comparisonGenTrip} presents the evaluation of two generator trip events: (1) G1 outage, resulting in a loss of $4\%$ of total system generation, and (2) G6 outage, resulting in a loss of $10.32\%$. 
% The accuracy of the estimation is measured by the absolute percentage error (APE), defined as \( APE = \left| \frac{\text{actual value} - \text{estimated value}}{\text{actual value}} \right| \times 100\% \). Results show a low estimation error, less than $10$ mHz and $0.25$ s. In terms of time performance, the average execution time for the predictions presented in this subsection is $0.01$ seconds. 

% The estimation i+s accurate in both scenarios, as indicated by the absolute percentage error (APE), defined as \( APE = \left| \frac{\text{actual value} - \text{estimated value}}{\text{actual value}} \right| \times 100\% \). In terms of time performance, the average execution time for the predictions presented in this subsection is 0.0100 seconds. 
\begin{figure*}[t]
    \centering
    % First row
    \begin{minipage}{0.23\textwidth}
        \centering
        \includegraphics[width=\textwidth]{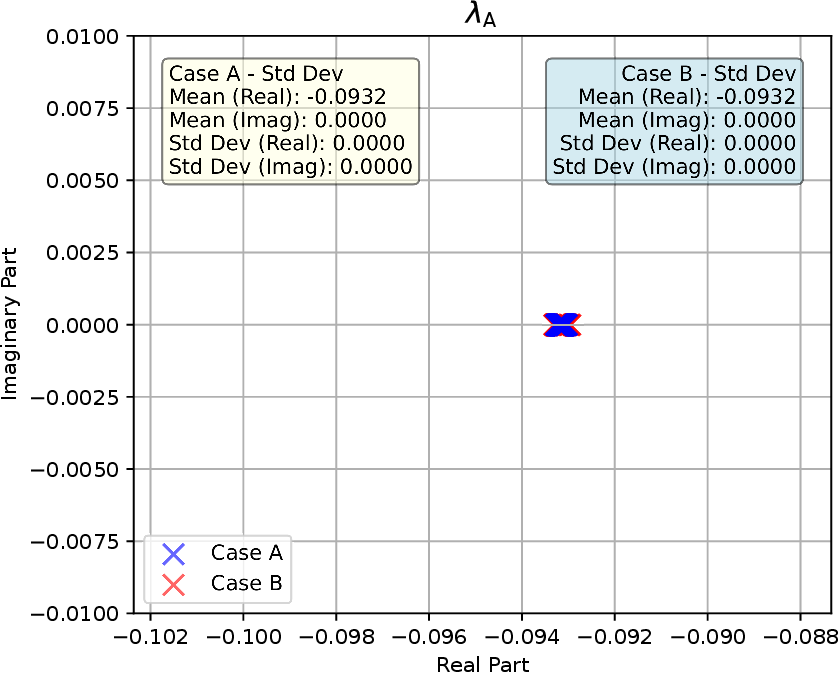}
        % \includegraphics[width=\textwidth]{figures/plot_lambda_A.pdf}
        % \textbf{(a)}
           % \scriptsize (a)
    \end{minipage}
    \hspace{0.5em}
    \begin{minipage}{0.23\textwidth}
        \centering
        \includegraphics[width=\textwidth]{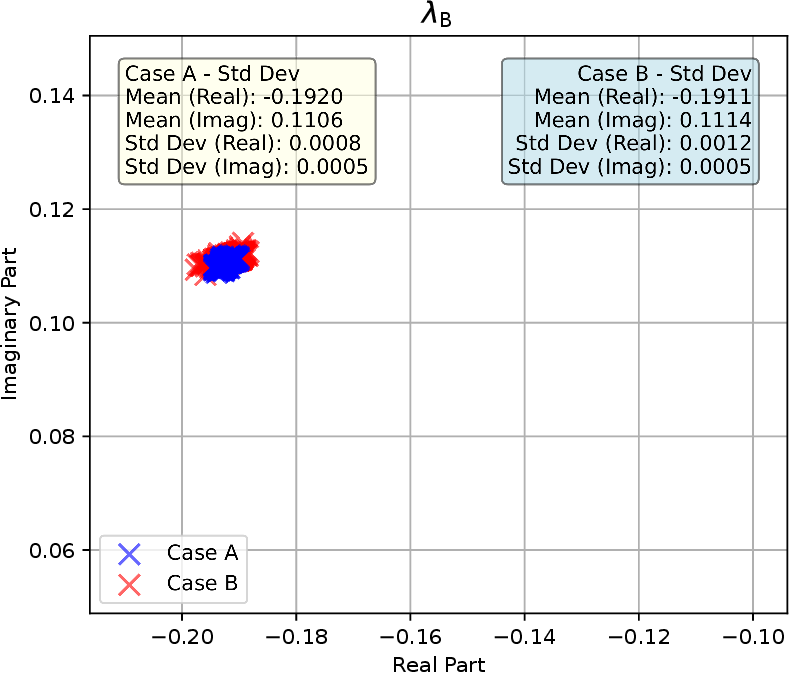}
        % \scriptsize (b)
    \end{minipage}
    \hspace{0.5em}
    \begin{minipage}{0.23\textwidth}
        \centering
        \includegraphics[width=\textwidth]{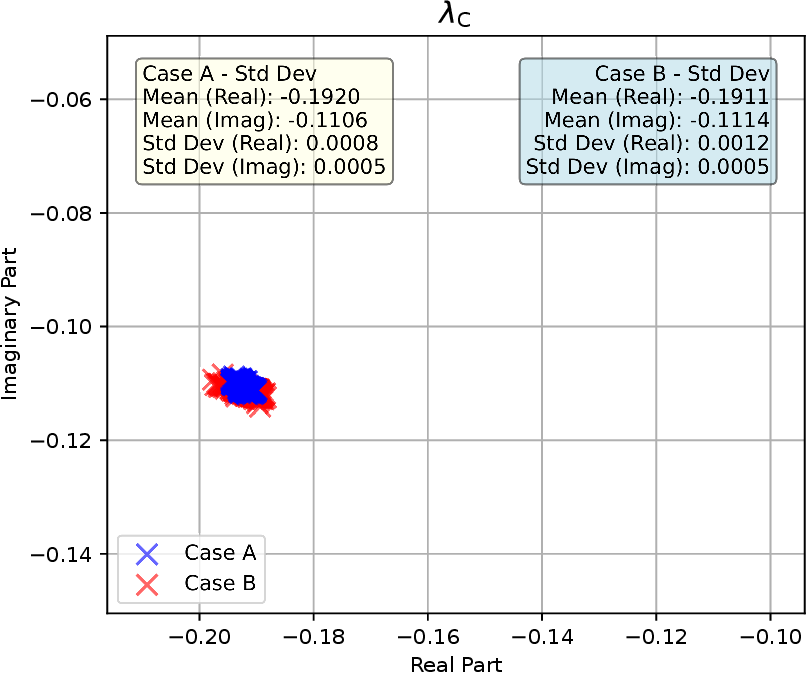}
        % \scriptsize (c)
    \end{minipage}
    \hspace{0.5em}
    \begin{minipage}{0.23\textwidth}
        \centering
         \includegraphics[width=\textwidth]{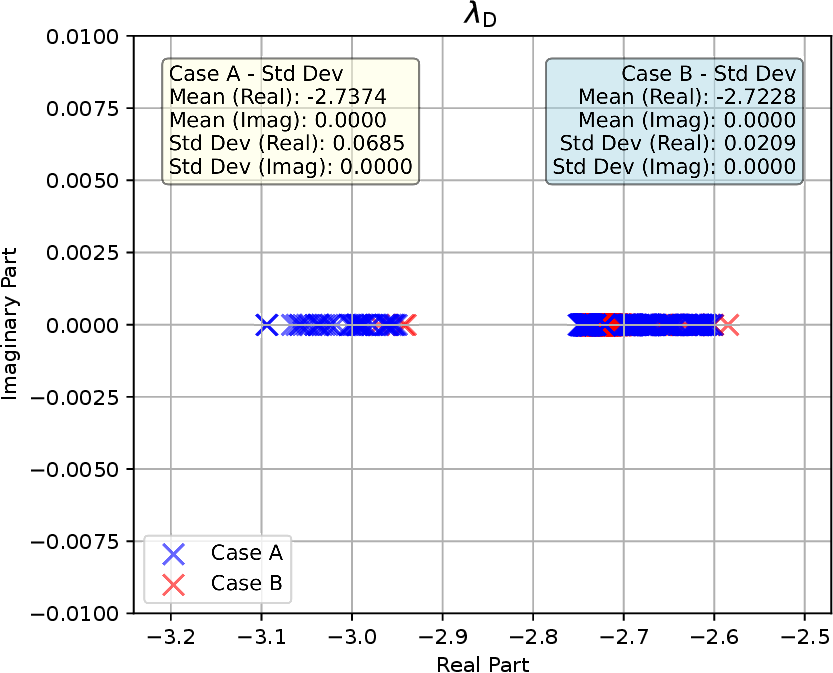}
        % \scriptsize (d)
    \end{minipage}
    
    % Second row
    \vspace{0.5em}    
    \caption{Frequency control modes sensitivity analysis in the 39-bus system. Case A: different loading levels, Case B: topology changes.}
    \label{fig:sensitivity_pf}
\end{figure*}

\begin{figure*}[t]
    \centering
    % First row
    \begin{minipage}{0.23\textwidth}
        \centering
        \includegraphics[width=\textwidth]{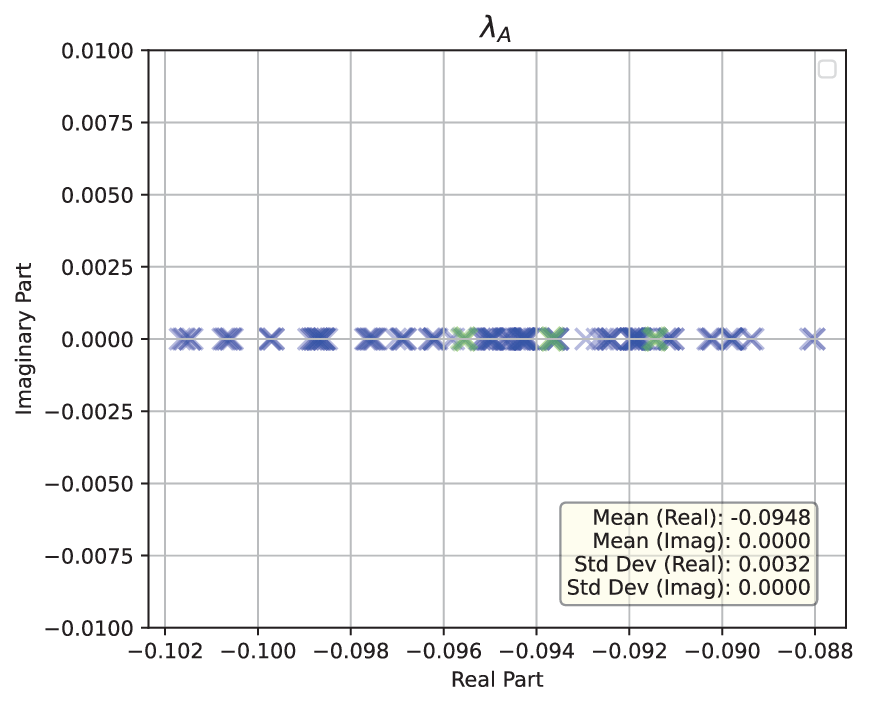}
        % \scriptsize (a)
    \end{minipage}
    \hspace{0.5em}
    \begin{minipage}{0.23\textwidth}
        \centering
        \includegraphics[width=\textwidth]{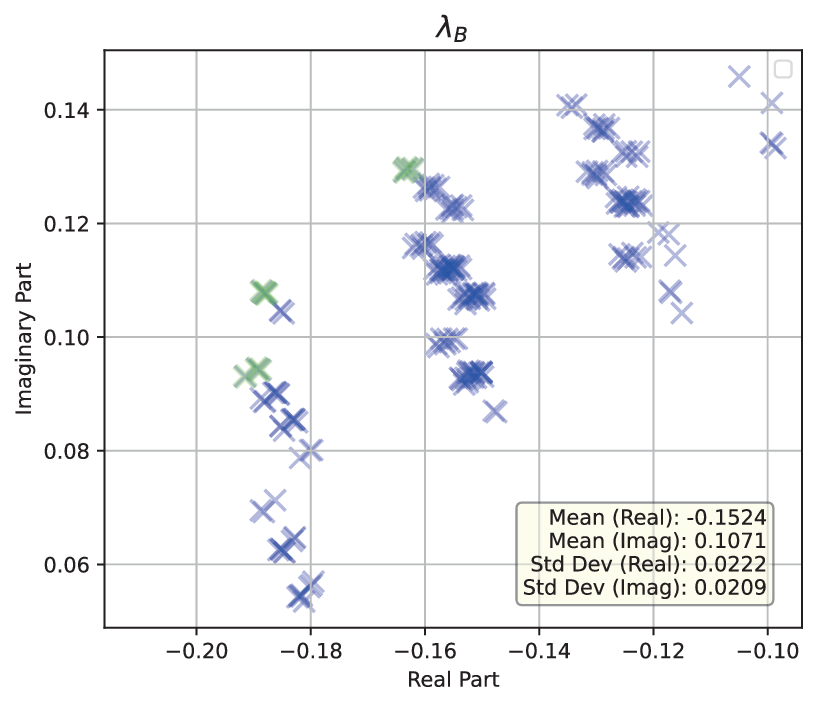}
        % \scriptsize (b)
    \end{minipage}
    \hspace{0.5em}
    \begin{minipage}{0.23\textwidth}
        \centering
        \includegraphics[width=\textwidth]{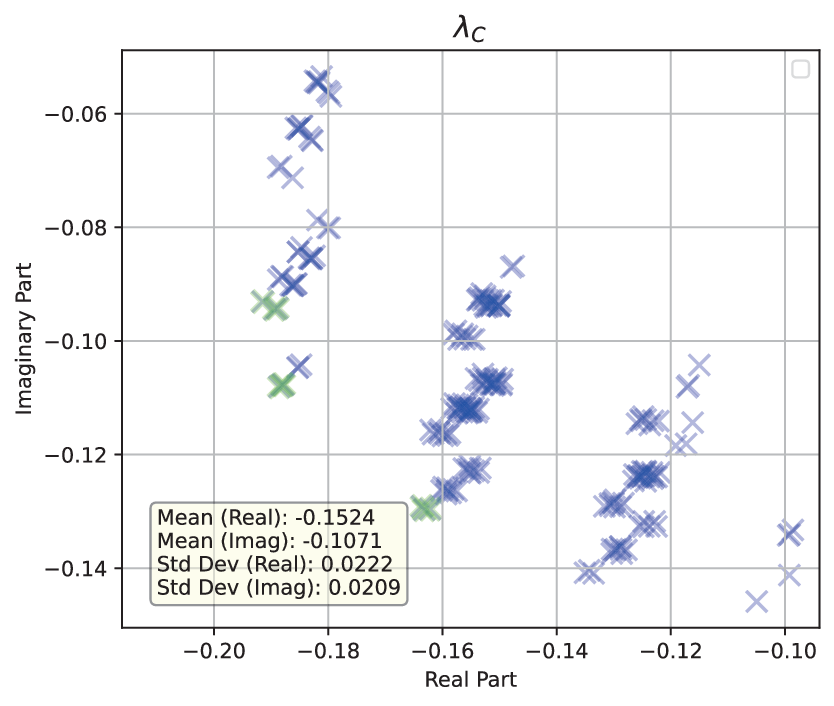}
        % \scriptsize (c)
    \end{minipage}
    \hspace{0.5em}
    \begin{minipage}{0.23\textwidth}
        \centering
         \includegraphics[width=\textwidth]{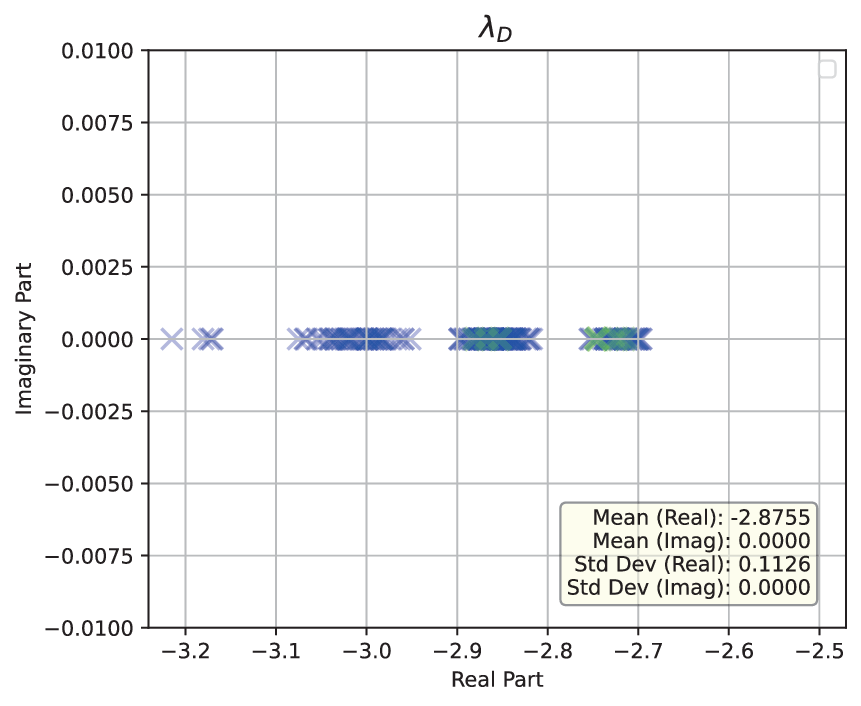}
        % \scriptsize (d)
    \end{minipage}
    
    % Second row
    \vspace{0.5em}    
    \caption{Frequency control modes sensitivity analysis in the 39-bus system for different unit commitments.}
    \label{fig:fullgrid}
\end{figure*}

\begin{table}[t]
    \centering
    \caption{Frequency nadir and time after G1 and G6 outages}
    \renewcommand{\arraystretch}{1.1}
    \setlength{\tabcolsep}{6pt} 
    \scriptsize
    \begin{tabular}{c|cc|cc}
        \toprule
        \textbf{Gen. Trip} & \multicolumn{2}{c|}{\textbf{G1 }} & \multicolumn{2}{c}{\textbf{G6 }} \\
        & Nadir & Time & Nadir & Time \\
        & (Hz) & (s) & (Hz) & (s) \\
        \midrule
        \textbf{Actual} & 59.489 & 9.6 & 58.961 & 9.4 \\
        \textbf{Estimation} & 59.493 & 9.35 & 58.971 & 9.31 \\
        \midrule
       \rowcolor{gray!20} \textbf{Absolute Error} & 0.004 & 0.25 & 0.01 & 0.09 \\
        \bottomrule
    \end{tabular}
    \label{tab:comparisonGenTrip}
\end{table}

Note that linearizing the system for every operating condition is unnecessary. Instead, it is sufficient to derive the modal information from a representative base case and extend it to other scenarios. This approach is justified because the system's modal information, corresponding to the frequency control modes, typically exhibits only minor variations across changes in loading levels, power flow patterns, and network topology. Even when the unit commitment changes, the modal information can often still reliably capture the system dynamics to a considerable extent. However, if the changes in the unit commitment are substantial, it becomes necessary to recompute the modal information to maintain accuracy.

A sensitivity analysis is conducted to expose the system to different scenarios, assessing both the variation of the frequency control modes and the accuracy of relying solely on base-case information for the estimation. 

% As previously shown, four modes are sufficient to perform the estimation. Figures \ref{fig:sensitivity_pf} and \ref{fig:fullgrid} illustrate the variation of these modes under different conditions.

\subsubsection{Case A – Different loading levels and power flows}
Five loading levels were analyzed, with dispatch randomly varied to reflect different power flow scenarios. A total of 2,363 simulations were performed: 1,103 with a 10\% load step at bus 15, and 1,260 with G1 outage. Fig.~\ref{fig:sensitivity_pf} shows the frequency control modes used in the estimation. Each subplot corresponds to one eigenvalue and its variation in the new steady state after the disturbance. The blue scatter plots (Case A) represent the full range of the eigenvalues across scenarios, with the mean and standard deviation included. The standard deviation for both the real and imaginary parts of the eigenvalues is on the order of $10^{-4}$, indicating minimal variation under different conditions. Nadir values were estimated using the base case eigenvalues and modal information. The accuracy of the estimation is measured by the mean absolute percentage error (MAPE), defined as $\text{MAPE} = \frac{1}{n} \sum_{i=1}^{n} \left| \frac{y_i - \hat{y}_i}{y_i} \right| \times 100\%$ where \( y_i \) is the actual value, \( \hat{y}_i \) is the estimated value, and \( n \) is the number of samples. The MAPE in frequency nadir and time was $0.068 \times 10^{-3}\%$ and $0.913\%$ for load step cases, and $0.068 \times 10^{-3}\%$ and $1.79\%$ for generator trip cases. These results confirm that using the base case modal data allows accurate prediction across different loading levels and power flow conditions.

\subsubsection{Case B – Different topologies and power flows}
A total of 2,095 scenarios were analyzed, including configurations where two or three transmission lines were intentionally taken out of service to represent alternative steady-state operating conditions. Of these, 995 scenarios involved a load step at bus 15, and 1,100 involved a generator G1 outage. Fig.~\ref {fig:sensitivity_pf} shows the modes used in the estimation. The red scatter plot (Case B) corresponds to the eigenvalues and their variation for this analysis. The standard deviation remains on the order of $10^{-3}$, indicating minimal deviation from the mean. In this analysis, the MAPE in frequency nadir and timing estimation for the load step scenarios is $0.036 \times 10^{-3}\%$ and $1.83\%$, respectively, and $0.075 \times 10^{-3}\%$ and $1.557\%$ for the generator trip. The use of base case modal information remains effective, as the modes show low sensitivity to changes in network topology.

\subsubsection{Case C – Different unit commitment}
This analysis evaluates the sensitivity of the modes to varying numbers and combinations of committed generators. Keeping the total demand constant, 354 scenarios were defined: 261 with a load step event (10\% at bus 15) and 93 with G1 outage. 

The blue scatter plots in Fig. \ref{fig:fullgrid} show the frequency control modes and their variation. Unlike the previous two analyses, this one reveals larger variability, particularly for $\lambda_A$, $\lambda_B$, and $\lambda_C$, which now exhibit a larger standard deviation on the order of $10^{-2}$—significantly affecting the period and damping of the modes. This outcome aligns with the fact that changes in generator participation affect system inertia and primary frequency support.

The MAPE in nadir and time estimation for the load step cases were $0.0099 \%$ and $14.57\%$, respectively, and $0.13\%$ and $13.598\%$ for the generator trip cases. Although nadir errors remain within acceptable bounds for load steps, time errors are significantly larger. In generator trip cases, the nadir error corresponds to approximately 0.07~Hz, which may be significant depending on the application. In Fig.\ref{fig:fullgrid}, green markers indicate eigenvalue variations under unit commitment scenarios modified by switching off a single generator from the base case. For this scenario, the use of the base case modal information keeps valid results within an acceptable error. The MAPE in nadir and time for the load step is reduced to $0.0017\%$ and $4.11\%$, and to $0.03 \%$ and $6.11\%$ for the generator trip. 

In summary, the modal information involved in the SFR and used for the estimation exhibits low sensitivity to variations in loading levels, power flows, and network topology. Additionally, its validity is maintained under small changes in unit commitment. These results indicate that linearizing the system at every possible operating condition is not required, which reduces computational effort while preserving estimation reliability. A complete summary of the results is presented in Table \ref{tab:sensitivity_summary}.

\begin{table*}[t]
\centering
\caption{Summary of the 39-bus system sensitivity analysis and estimation errors}
\label{tab:sensitivity_summary}
\begin{tabular}{c|c|c|c|c|c}
\toprule
\textbf{Scenario} & \textbf{Disturbance} & \textbf{Cases} & \textbf{Stand. Dev. Order} & \textbf{MAPE Nadir (\%)} & \textbf{MAPE Time (\%)} \\ \hline
\multirow{2}{*}{Loading Levels and Power Flows} & Load Step & 1,103 & $10^{-4}$ & $0.068\times 10^{-3}$ & 0.913 \\ \cline{2-6}
 & Generator Trip & 1,260 & $10^{-4}$ & $0.068\times 10^{-3}$ & 1.79 \\ \hline
\multirow{2}{*}{Topologies and Power Flows} & Load Step & 995 & $10^{-3}$ & $0.036\times 10^{-3}$ & 1.83 \\ \cline{2-6}
 & Generator Trip & 1,100 & $10^{-3}$ & $0.075\times 10^{-3}$ & 1.557 \\ \hline
\multirow{2}{*}{Generator Participation } & Load Step & 261 & $10^{-2}$ & $0.0099$ & 14.57 \\ \cline{2-6}
 & Generator Trip & 93 & $10^{-2}$ & $0.13$ & 13.598 \\ \hline
\multirow{2}{*}{Generator Participation (Removing 1 machine)} & Load Step & 9 & $10^{-3}$ & $0.0017$ & 4.11 \\ \cline{2-6}
 & Generator Trip & 9 & $10^{-3}$ & $0.03$ & 6.11 \\ 
 \bottomrule
\end{tabular}
\end{table*}

\subsection{El Salvador's transmission planning (ESTP) system}
The method is validated on the ESTP network, which includes 145 buses, 64 transmission lines, and 28 power plants (thermal, hydro, photovoltaic, and wind), operating at voltage levels from 23 kV to 230 kV. The system has an installed capacity of 2,322.6 MW and a peak demand of 1,131 MW. As part of the SIEPAC (Central American Electrical Interconnection System), it also features two 230 kV interconnections with Guatemala and Honduras. Further details and the full database are available on the national system operator’s website \cite{unidad2025}. A key scenario occurs when these interconnections are lost, isolating the system from the rest of SIEPAC. Under such conditions, frequency deviations become more pronounced due to increased sensitivity. 

The system is evaluated under two dry-season contingencies at minimum demand (631.45 MW) and isolated operation, using 20 generating units (steam, gas, and hydro), yielding a 259-state-variable model. The scenarios involve 15 MW and 50 MW load steps at Pedregal and Ateos—major load centers with high distributed PV penetration. Linearization and dynamic simulations are conducted in PowerFactory.

\begin{figure}[t]
    \centering
    \hspace{-0.9cm} % Adjust this value as needed
    \includegraphics[width=0.53\textwidth]{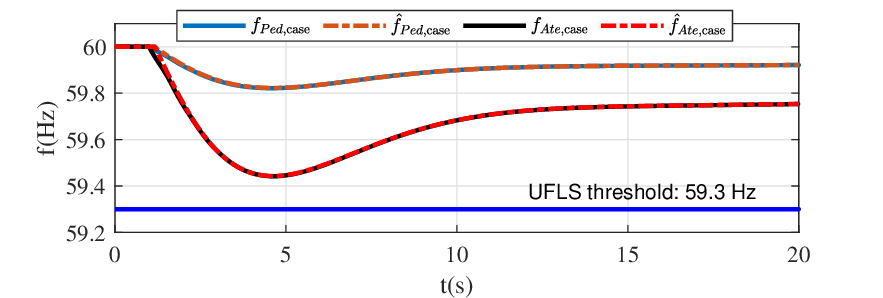}
    \caption{ESTP frequency response comparison: actual vs estimated.}
    \label{fig:comparison_sites}
\end{figure}

For estimation, 9 of the 259 eigenvalues are selected—3 real and 3 complex conjugate pairs. Fig.~\ref{fig:comparison_sites} shows the predicted frequency responses compared to PowerFactory simulations, and Table~\ref{tab:comparisonESA} reports nadirs, time, and absolute errors. The Ateos event proves most critical, with frequency dropping to 59.441 Hz—near the UFLS threshold—closely matched by the prediction. The method achieves high accuracy with a fast average execution time of 0.0205 seconds, even when the system has more state variables and the frequency trajectory is computed by a large set of dynamics.
\begin{table}[t]
    \centering
    \caption{ESTP frquency nadir and time estimation}
    \renewcommand{\arraystretch}{1.1}
    \setlength{\tabcolsep}{6pt} 
    \scriptsize
    \begin{tabular}{c|cc|cc}
        \toprule
        \textbf{Location} & \multicolumn{2}{c|}{\textbf{Pedregal}} & \multicolumn{2}{c}{\textbf{Ateos}} \\
        & Nadir & Time & Nadir & Time \\
        & (Hz) & (s) & (Hz) & (s) \\
        \midrule
        \textbf{Actual} & 59.821 & 4.60 & 59.441 & 4.64 \\
        \textbf{Estimation} & 59.822 & 4.62 & 59.443 & 4.62 \\
        \midrule
       \rowcolor{gray!20} \textbf{Absolute Error} & 0.001 & 0.02 & 0.002 & 0.02 \\
        \bottomrule
    \end{tabular}
    \label{tab:comparisonESA}
\end{table}

These findings suggest that the approach can extend to more complex systems and may be compatible with commonly used software tools. Furthermore, results show that even when the number of state variables considerably increases, a small subset of dynamics is sufficient to accurately compute the predictions.

\section{Inverter-Based Resources Considerations}
To evaluate the impact of IBRs on SFR dynamics, frequency nadir and time estimation, the Renewable Energy Generation Converter-B (REGC-B) model is used along with the controllers presented in \cite{ramasubramanian2016converter}. Detailed information on the converter, controllers, and their DAE formulations is provided in \cite{MartinezLizana2025}, while parameter values are listed in Appendix \ref{IBR_parameters}. The analysis is conducted using the New England 10-machine, 39-bus test system.

Two control modes are considered for IBRs: active/reactive power control (PQ-control) and frequency/voltage control (FV-control). To assess their impact on the proposed SFR and nadir estimation method, we analyze how each control mode influences estimation accuracy. Specifically, we examine the shifts of estimated modes in the complex plane and the participation of IBR state variables. Results indicate that, regardless of IBR participation or control mode, system dynamics remain dominated by SGs, with minimal modal variation. These findings support the method's robustness across different control strategies and levels of IBR integration.

\subsection{PQ-control operation}
This operating mode regulates the current injected at the interconnection point to maintain the active and reactive power setpoints, regardless of system disturbances. Consequently, it does not participate in frequency support. The New England system is evaluated under different scenarios of IBR penetration. The total system demand remains constant, with SGs progressively replaced by IBRs to increase the penetration level. Nadir estimation is assessed under the same disturbance across all scenarios—a 20\% step increase in load at bus 7. Table~\ref{tab:ibr_penetration} presents the IBR penetration percentages, system inertia values, and both actual and predicted nadir and nadir timing results for all scenarios. The results show an increase in nadir deviation with higher IBR penetration and reduced system inertia. However, the estimation remains accurate across all scenarios. 
\begin{table*}[ht]
\centering
\caption{Impact of IBR penetration levels and operating modes (PQ and FV control)\\in the 39-bus system frequency nadir and time estimation}
\begin{tabular}{c|ccc|cccc|cccc}
\toprule
\multirow{2}{*}{Scenario} & \multicolumn{3}{c|}{\textbf{Systems data}} & \multicolumn{4}{c|}{\textbf{PQ-control operation}} & \multicolumn{4}{c}{\textbf{FV-control operation}} \\
 & \begin{tabular}[c]{@{}c@{}}IBR Gen.\\ (\%)\end{tabular} & \begin{tabular}[c]{@{}c@{}}SG Gen.\\ (\%)\end{tabular} & \begin{tabular}[c]{@{}c@{}}System\\ Inertia (s)\end{tabular} & \begin{tabular}[c]{@{}c@{}}Actual\\ Nadir\end{tabular} & \begin{tabular}[c]{@{}c@{}}Pred.\\ Nadir\end{tabular} & \begin{tabular}[c]{@{}c@{}}Actual\\ Time (s)\end{tabular} & \begin{tabular}[c]{@{}c@{}}Pred.\\ Time (s)\end{tabular} & \begin{tabular}[c]{@{}c@{}}Actual\\ Nadir\end{tabular} & \begin{tabular}[c]{@{}c@{}}Pred.\\ Nadir\end{tabular} & \begin{tabular}[c]{@{}c@{}}Actual\\ Time (s)\end{tabular} & \begin{tabular}[c]{@{}c@{}}Pred.\\ Time (s)\end{tabular} \\
 \hline
1 & 0.00 & 100.00 & 313.08 & 59.8610 & 59.8634 & 9.0963 & 9.0371 & 59.8610 & 59.8634 & 9.0963 & 9.0371 \\
2 & 3.96 & 96.03  & 296.28 & 59.8501 & 59.8530 & 9.3552 & 9.4461 & 59.8636 & 59.8661 & 9.0752 & 8.9781 \\
3 & 14.29 & 85.70 & 281.96 & 59.8207 & 59.8246 & 10.1678 & 10.1948 & 59.8548 & 59.8576 & 9.0746 & 9.1246 \\
4 & 22.35 & 77.64 & 271.56 & 59.8050 & 59.8091 & 10.7587 & 10.7672 & 59.8588 & 59.8616 & 9.2466 & 9.0572 \\
5 & 31.12 & 68.87 & 257.64 & 59.7545 & 59.7592 & 11.9002 & 11.8802 & 59.8497 & 59.8523 & 8.9441 & 9.1855 \\
6 & 44.60 & 55.39 & 243.84 & 59.6729 & 59.6779 & 13.4693 & 13.4878 & 59.8394 & 59.8421 & 9.4405 & 9.3618 \\
\bottomrule
\end{tabular}
\label{tab:ibr_penetration}
\end{table*}

\begin{figure}[t]
    \centering
    \includegraphics[width=0.4\textwidth]{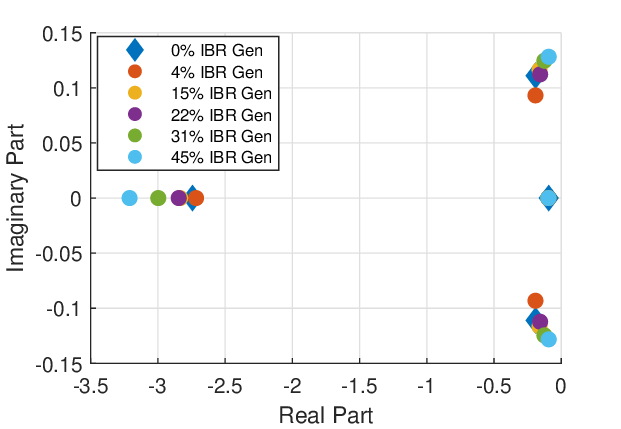} % 80% of the text width
    \caption{Frequency control modes under varying penetration levels of IBR PQ-control for the 39-bus system.}
    \label{fig:pq_eigenvalues}
\end{figure}

To establish a clear reference point, the base case—corresponding to $0\%$ IBR generation—is defined using the same operating conditions discussed in Section~\ref{subsec:NE}. The SFR and nadir under various IBR penetration levels are primarily governed by four dominant modes, which are identified through participation factor-guided mode selection. These modes, along with their evolution as IBR contribution increases, are illustrated in Fig.~\ref{fig:pq_eigenvalues}. Notably, despite the growing share of IBRs in the generation mix, the dominant dynamic behavior of the SFR remains largely unchanged, exhibiting only minor variations. Furthermore, all participation factors corresponding to these dominant eigenvalues and associated with IBR state variables consistently remain below the threshold $\epsilon = 10^{-4}$, indicating their negligible direct influence on the frequency control modes.

\subsection{FV-control operation}
In this case, the IBR regulates the voltage at its interconnection point and provides frequency support to the grid through droop control. This enables it to participate in grid support by responding to disturbances. Since IBRs participate in frequency regulation, their droop characteristics must be considered in the initial condition computation described in Section~\ref{subsection: init con}. In particular, the droop behavior of FV-control IBRs must be included in Eq.~\ref{eq:Post_freq} by accounting for their speed regulation, such that
$\Delta \omega = -\Delta P \big/ \left( \textstyle\sum_{z \in \mathcal{Z}} \frac{1}{R_{D,z}} + \sum_{u \in \mathcal{U}} \frac{1}{R_{p,u}} \right)$,
where $\mathcal{Z}$ and $\mathcal{U}$ denote the sets of synchronous generators and IBRs, respectively. This modification affects the computation of the post-disturbance steady-state frequency. 

\begin{figure}[h]
    \centering
    \includegraphics[width=0.4\textwidth]{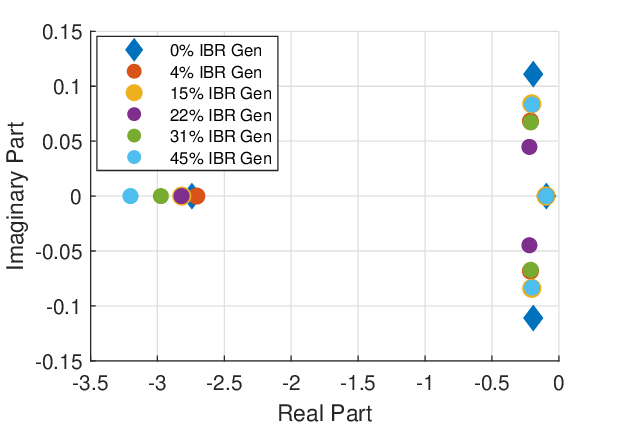} % 80% of the text width
    \caption{Frequency control modes under varying penetration levels of IBR FV-control for the 39-bus system.}
    \label{fig:fv_eigenvalues}
\end{figure}

The SFR and nadir in this case are again determined by the same four dominant eigenvalues and their associated modal structures, as illustrated in Fig.\ref{fig:fv_eigenvalues}. Although the total number of state variables and eigenvalues changes across different IBR penetration, no additional modes are required. The key eigenvalues governing the SFR show only minor variations, confirming the persistence of the dominant dynamics identified previously. As shown in Table\ref{tab:ibr_penetration}, increased IBR participation in FV-control mode leads to reduced frequency deviation, while the proposed prediction method continues to yield accurate results even under these altered conditions, reinforcing its effectiveness across diverse operational scenarios.
% The system frequency response and nadir are obtained using the same four eigenvalues and their associated modal information (Fig.~\ref{fig:fv_eigenvalues}). Although the eigenvalues vary, no new ones are needed. While the number of state variables and eigenvalues changes with different IBR or SG scenarios, the key eigenvalues and their modal characteristics defining the SFR remain largely unchanged, with only minor variations, as illustrated in the complex plots throughout this work.

% Table~\ref{tab:ibr_penetration} shows that participation of IBRs in frequency regulation reduces the magnitude of frequency deviation and that the proposed prediction approach provides good estimates even in this scenario. 

In the participation factor analysis, IBR states associated with droop control exceed the threshold $\epsilon$ in one of the four modes, but remains below $\epsilon$ in the others. This observation suggests that, even when IBRs actively contribute to frequency regulation, the system's frequency response is still predominantly governed by the dynamics and frequency control of SGs. It is reasonable to expect that modifications to the controllers or models of the converters would influence the existing modes or give rise to new ones. However, the system response can be reliably reconstructed by focusing on the subset of dominant modes, as demonstrated by the method proposed in this work. 

\section{Conclusion}
This work introduces a modal-based analytical methodology for predicting system frequency response and estimating the frequency nadir, grounded in the system's eigenstructure. By performing modal decomposition and identifying dominant modes, a closed-form expression of the frequency trajectory is derived, enabling accurate and computationally efficient assessments. The results show that these dominant modes exhibit low sensitivity to parameter variations, ensuring robust performance across a wide range of operating conditions, including those with high penetration of inverter-based resources. Validation on two benchmark systems and the Salvadoran transmission network confirms the method’s accuracy, scalability, and practical applicability. This shift from average-based representations to a mode-driven analysis underscores the value of leveraging system eigenstructure to capture essential dynamics. Future work will focus on integrating this framework into real-time dynamic security assessment and improving planning efficiency.
% \clearpage

\appendices
\section{Coefficients for nadir time estimation}
\vspace{-0.4cm}
\begin{align}
    \psi &= \beta \tau+ \theta + \phi, \ \quad K_1 = R e^{\alpha \tau}, \quad K_2 = \sqrt{\alpha^{2}+\beta^{2}}\\
    a_1 &= K_1 \left( \frac{(\alpha^2 - \beta^2)}{2} \cos(\psi) - \alpha \beta \sin(\psi) \right) \\
    a_2 &= K_1 \left( K_2 \cos(\psi+\phi) - \right. \nonumber \\
    & \quad \left. ((\alpha^2 -\beta^2)\cos(\psi)-2\alpha\beta\sin(\psi))\tau \right)\\
    a_3 &= K_1 \left(\cos(\psi)-K_2\cos(\psi+\phi)\tau + \right. \nonumber
    \\
    & \quad \left. ( \frac{(\alpha^2 - \beta^2)}{2} \cos(\psi) - \alpha \beta \sin(\psi))\tau^2  \right)  \\ 
    a_4 &= \frac{\lambda_{r}^3 e^{\lambda_{r} t_0} \gamma_{r}}{2}, \ \  a_5 = \lambda_{r}^2 e^{\lambda_{r} \tau} \left(1 - \lambda_{r} \tau \right) \gamma_{r}  \\
    a_6 &= \lambda_{r} e^{\lambda_{r} \tau}\left(1 - \lambda_{r} \tau + \frac{\lambda_{r}^2 \tau^2}{2} \right)\gamma_{r}
\end{align}
% \begin{align*}
%     a_3 &= K_1 \left(\cos(\psi)-K_2\cos(\psi+\phi)\tau + \right. 
%     \\
%     & \quad \left. ( \frac{(\alpha^2 - \beta^2)}{2} \cos(\psi) - \alpha \beta \sin(\psi))\tau^2  \right)  \\ 
%     a_4 &= \frac{\lambda_{r}^3 e^{\lambda_{r} t_0} \gamma_{r}}{2}, \ \  a_5 = \lambda_{r}^2 e^{\lambda_{r} \tau} \left(1 - \lambda_{r} \tau \right) \gamma_{r},  \\
%     a_6 &= \lambda_{r} e^{\lambda_{r} \tau}\left(1 - \lambda_{r} \tau + \frac{\lambda_{r}^2 \tau^2}{2} \right)\gamma_{r}.
% \end{align*}

\section{Governor Parameters for Benchmark Models}
\vspace{-0.15in}
\label{tab:governor_params}
\begin{table}[H]
    \centering
    \caption{WSCC 9-bus test System}
    \begin{tabular}{c c c c c c c c c c c}
        \hline
        Bus & $T_1$ & $T_2$ & $T_3$ & $T_4$ & $K_1$ & $A_t$ & $K_t$ & $D_t$ \\
        \hline
        1 & 0.25 & 2.50  & 9.00  & 0.00 & 20.00 & 0 & 0.0 & 0.2 \\ % TGOV1
        2 & 0.30 & 5.00  & 12.00 & 0.10 & 30.00 & 0 & 0.0 & 0.0 \\ % IEESGO
        3 & 0.25 & 0.25  & 2.50  & 0.00 & 25.00 & 1 & 2.5 & 0.2 \\ % GAST
        \hline
        \vspace{0.005in}
    \end{tabular}
    % \vspace{0.85cm}
    \parbox{0.85\linewidth}{\footnotesize Note: IEESGO parameters for generator at bus 2: $K_{2}=K_{3}=0$.}
\end{table}

\begin{table}[H]
    \centering
    \caption{New England 39-bus test System}
    \resizebox{\columnwidth}{!}{%
    \begin{tabular}{c c c c c c c c c c}
        \hline
        Bus & Type & $T_1$ & $T_2$ & $T_3$ & $T_4$ & $K_1$ & $D_t$ & $A_t$ & $K_t$ \\
        \hline
        1  & 1  & 0.25 & 2.50  & 9.00  & 0.00 & 27.00 & 0.25 & 0.00 & 0.00 \\ % TGOV1
        2  & 2 & 0.30 & 5.00  & 12.00 & 0.10 & 23.00 & 0.00 & 0.00 & 0.00 \\ % IEESGO
        3  & 3   & 0.25 & 0.25  & 2.50  & 0.00 & 33.00 & 0.25 & 8.00 & 2.50 \\ % GAST
        4  & 1  & 0.25 & 2.50  & 9.00  & 0.00 & 27.00 & 0.25 & 0.00 & 0.00 \\ % TGOV1
        5  & 2 & 0.30 & 5.00  & 12.00 & 0.10 & 23.00 & 0.00 & 0.00 & 0.00 \\ % IEESGO
        6  & 3   & 0.25 & 0.25  & 2.50  & 0.00 & 33.00 & 0.25 & 7.00 & 2.50 \\ % GAST
        7  & 1  & 0.25 & 2.50  & 9.00  & 0.00 & 27.00 & 0.25 & 0.00 & 0.00 \\ % TGOV1
        8  & 2 & 0.30 & 5.00  & 12.00 & 0.10 & 23.00 & 0.00 & 0.00 & 0.00 \\ % IEESGO
        9  & 3   & 0.25 & 0.25  & 2.50  & 0.00 & 33.00 & 0.25 & 9.00 & 2.50 \\ % GAST
        10 & 1 & 0.30 & 5.00  & 12.00 & 0.10 & 25.00 & 0.00 & 0.00 & 0.00 \\ % IEESGO
        \hline
        \vspace{0.005in}
    \end{tabular}
    }
    \parbox{\columnwidth}{\footnotesize Note: Type 1 corresponds to TGOV1, Type 2 corresponds to IEESGO, and Type 3 corresponds to GAST.}
    \label{tab:governor_params_39bus}
\end{table}

\section{ IBR parameters}
\label{IBR_parameters}
\begin{table}[H]
\centering
\caption{REGC-B dynamic parameters}
\begin{tabular}{cccccccc}
\hline
$T_Q$ & $T_D$ & $R_f$ & $X_f$ & $T_{eq}$ & $T_{ed}$ \\
\hline
 0.01 & 0.01 & 0.0040 & 0.05 & 0.01 & 0.01 \\
\hline
\end{tabular}
\label{tab:dyn_ibr}
\end{table}

\begin{table}[H]
\centering
\caption{IBR control parameters}
 \resizebox{\columnwidth}{!}{%
\begin{tabular}{cccccccccccccccc}
\hline
 $K_{iq}$ & $T_{Gqv}$ & $K_{ip}$ & $T_{Gpv}$ & $T_r$ & $R_q$ & $K_i$ & $K_p$ & $T_{frq}$ & $R_p$ \\
\hline
5 & 0.01 & 5 & 0.01 &  0.02 & 0.00 & 20.0 & 4.00 & 0.01 & 0.05 \\
\hline
\vspace{0.005in}
\end{tabular}
}
\parbox{\columnwidth}{\footnotesize Note: For PQ-control operation $T_r$, $K_i$, $K_p$, $T_{frq}$, $R_p$ parameters are set to 0.00.}
\label{tab:pq_control}
\end{table}

\bibliographystyle{IEEEtran}
% \small
\bibliography{Biblio}

% \textcolor{red}{Aun falta definir $a_1$, $a_2$ y $a_3$.}
% \begin{align*}
% \lambda_{l} = \alpha_{l} + j \beta_{l} \ \ \ r_{k,l} = \sqrt{e_{k,l}^2 + f_{k,l}^2} \ \ \ \theta_k = \tan^{-1} \left( \frac{f_{k,l}}{e_{k,l}} \right) \\
% a_{l}+j b_{l}=w^{T}_{l}\Delta x_{0} \quad \quad c_{k,l} + j d_{k,l} = v_{l,k} \ \ \ \ \ \ \ \ \ \  \\
% e_{k,l} = a_l \cdot c_{k,l} - b_l \cdot d_{k,l} \ \ \ f_{k,l} = a_l \cdot d_{k,l} + b_l \cdot c_{k,l} \ \ \ \ 
% \end{align*}

%%%%%%%%%%%%%%%%%%%%%%%%%%%%%%%%%%%%%%%%%%%%%%%%%%%%%%%%%%%%%%%%%%%%%%%%%%%%%%%%

% \bibliography{bibliog}
\end{document}